\documentclass[12pt,a4paper]{article}
\usepackage{amsmath}
\usepackage{epsfig}
\usepackage{array}
\usepackage{float}
\usepackage{lscape,graphicx} 
\usepackage{graphics}
\usepackage{amssymb}
\usepackage{color}
\usepackage{multirow}
\usepackage{advdate}
\usepackage{datenumber}
\usepackage{ mathrsfs }
\usepackage[utf8]{inputenc}

\newcommand{\gev}{\mbox{${\rm GeV}$}}

\newcommand{\bea}{\begin{equation}\begin{array}{c}}
\newcommand{\eea}{\end{array}\end{equation}}
\newcommand{\ea}{\end{array}}

\newcommand{\beq}{\begin{equation}}
\newcommand{\eeq}{\end{equation}}
\newcommand{\bad}{\begin{array}{ccc}}

\newcommand{\ba}{\begin{array}{c}}

\newcommand{\sigmav}{\langle\sigma v\rangle}
\newcommand{\ubl}{U(1)_{B-L}}
\newcommand{\mzbl}{M_{Z_{BL}}}
\newcommand{\gbl}{g_{BL}}
\newcommand{\ndm}{n_{DM}}
\newcommand{\zbl}{{Z_{BL}}}
\newcommand{\mdm}{M_{\phi_{DM}}}
\newcommand{\tev}{\mathrm{TeV}}
\newcommand{\sip}{\sigma_{SI}}
\newcommand{\mni}{M_{N_i}}
\newcommand{\mhtwo}{M_{H_2}}
\newcommand{\ldm}{\lambda_{DM}}
\newcommand{\phidm}{\phi_{DM}}

\newcommand{\sbl}{S_{BL}}
\newcommand{\vbl}{v_{BL}}

\begin{document}

\title{
{\bf Scalar dark matter in the $\mathbf{B-L}$ model}}
\author{Werner Rodejohann$\footnote{werner.rodejohann@mpi-hd.mpg.de}$~  and Carlos E. Yaguna$\footnote{carlos.yaguna@mpi-hd.mpg.de}$\\[7mm]
\it   Max-Planck-Institute f\"ur Kernphysik,\\
\it \small Saupfercheckweg 1, 69117 Heidelberg, Germany
}
\date{}
\maketitle
\thispagestyle{empty}
\begin{abstract}
The $\ubl$ extension of the Standard Model requires the existence of right-handed neutrinos and naturally realizes the seesaw mechanism of neutrino mass generation.   We study the possibility of explaining the dark matter in this model with an additional scalar field, $\phidm$, that is a singlet of the Standard Model but charged under $\ubl$. An advantage of this scenario is that the stability of $\phidm$ can be guaranteed by appropriately choosing its $B-L$ charge, without the need of an extra ad hoc discrete symmetry. We investigate in detail the dark matter phenomenology of this model. We show that  the observed dark matter density can be obtained via gauge or scalar interactions, and that semi-annihilations could play an important role in the latter case. The regions consistent with the dark matter density are determined in each instance and the prospects for detection in future experiments are analyzed. If dark matter annihilations are controlled  by the $B-L$ gauge interaction, the mass of the dark matter particle should lie below 5 TeV and  its direct detection cross section can be easily probed by XENON1T; if instead they are controlled by scalar interactions, the dark matter mass can be much larger and the detection prospects are less certain.  Finally, we show that this scenario can be readily extended to accommodate multiple dark matter particles. 
\end{abstract}
\newpage
\section{Introduction}
About $27\%$ of the energy-density of the Universe consists of an exotic form of matter commonly known as dark matter \cite{Ade:2013zuv}. Throughout the years, many different dark matter candidates have been considered  in the literature but so far no evidence has been found for any single one. Among them, Weakly Interacting Massive Particles (WIMPs) are particularly attractive for several reasons.   First, particles with masses between  the GeV and the  TeV scale and with  weak-strength interactions appear in many extensions of the Standard Model including the MSSM and UED models. Second, such particles  naturally achieve, via a freeze-out  in the early Universe, a  relic density comparable to the observed value of the dark matter density --the so-called WIMP miracle. Third, WIMPs can be probed with current experiments in several ways, including direct \cite{Cushman:2013zza} and indirect \cite{Buckley:2013bha} dark matter searches as well as collider searches at the LHC \cite{ATLAS:2012zim}. In fact, we seem to be living through a very special time, the epoch in which the WIMP paradigm, which  has dominated dark matter research for many years, is either going to be excluded or confirmed \cite{Bertone:2010at}.

A problem common to many WIMP  models is the explanation of the dark matter stability.  It is indeed strange to have such a heavy particle ($M\sim \mathrm{TeV}$) to  be stable. Generally, dark matter stability is guaranteed by some  discrete symmetry introduced exclusively for that purpose, usually a $Z_2$. But that use of discrete symmetries is questionable not only due to its lack of motivation but also  because they are expected to be broken by gravitational effects at the Planck scale, inducing dark matter decay and likely destroying the feasibility of such models. That is why it is often implicitly assumed that such discrete symmetries are  actually the remnants of additional gauge \cite{Batell:2010bp} or flavor \cite{Sierra:2014kua} symmetries present at a higher scale, thereby delegating the problem to a framework larger than the model under consideration. It is not always clear, however, whether those assumptions  are actually viable  in such models.  A better approach may be  to find complete models where the dark matter is automatically stable due to the model's structure and the quantum numbers of the dark matter field, without additional discrete symmetries \cite{Cirelli:2005uq,Hambye:2008bq}. Several models of this type have been studied in the literature \cite{Frigerio:2009wf, Duerr:2013lka, Lindner:2013awa}, and they generally feature extended gauge sectors.

In this paper, we consider a minimal extension of the Standard Model by a $\ubl$ gauge group. This extension is particularly appealing because it is remarkably simple, it is anomaly free once a right-handed neutrino per generation is introduced, and it naturally realizes the seesaw mechanism of neutrino mass generation. In addition, the $\ubl$ model has a rich collider phenomenology \cite{Basso:2008iv} due to the presence of another neutral gauge boson,  $\zbl$, that couples to quarks and leptons.  This minimal setup, however, cannot  account for the dark matter. To do so one needs either to modify the symmetry or increase the particle content. The former approach was applied in several works \cite{Okada:2010wd,Okada:2012sg,Basak:2013cga,Sanchez-Vega:2014rka}, where an additional $Z_2$ symmetry was used to distinguish one of the particles, the would-be dark matter candidate. More natural seems  the idea of increasing the particle content with an additional fermion or scalar field charged under $B-L$, with this charge chosen in such a way to guarantee its stability. For fermions, a vector-like pair is needed to not spoil anomaly cancellation, and the resulting scenario is very predictive, as emphasized recently in \cite{Duerr:2015wfa}.  Scalars, on the other hand, are not constrained by anomaly cancellation, so one more field is enough and, having both scalar and gauge interactions, typically give rise to a richer phenomenology. To our knowledge, scalar dark matter in this setup has been previously studied only in \cite{Guo:2015lxa}, which exclusively considered  the scale invariant version of the $B-L$ model.  Here we want to study instead  scalar dark matter within the general $B-L$ extension of the Standard Model.  Specifically, we will analyze the different ways in which the relic density constraint can be satisfied, including resonant and non-resonant annihilations as well as semi-annihilations;   we will determine the viable parameter space in each case; and we will investigate  the detection prospects in future experiments. As we will show, when the dark matter density is determined by gauge-mediated interactions, the resulting framework is quite predictive and can be entirely probed by XENON1T. Finally, we also point out that  multi-component dark matter can easily be accommodated within this scenario.

The rest of the paper is organized as follows. In the next section we introduce our notation and describe in some detail the model. Our main results are presented in Section \ref{sec:dm}. In it we numerically study the dark matter phenomenology of the model according to the interactions (gauge or scalar) and processes (annihilations or semi-annihilations) that set the relic density, and we analyze the detection prospects in each case. In Section \ref{sec:multi} we qualitatively discuss  how to extrapolate our results to scenarios with multiple dark matter particles. Finally, we summarize our findings and draw our  conclusions  in Section \ref{sec:con}.

\section{The model}
\label{sec:mod}
Under $\ubl$ the Standard Model quarks and leptons have  charge $1/3$ and $-1$ respectively. As a result, $B-L$ is an anomalous symmetry and it is necessary to add additional fermions to consistently gauge it. In this paper, we consider a model based on the gauge symmetry $SU(3)\times SU(2)\times U(1)_Y\times\ubl$ and containing  three generations of right-handed neutrinos ($N_i$, $i=1,2,3$) to cancel the gauge anomalies, and a scalar, $S_{BL}$, singlet of the Standard Model but charged under $\ubl$,  that spontaneously breaks the $B-L$ symmetry. We will assign charge $+2$ to $S_{BL}$ so that the $N_i$ acquire Majorana masses upon the breaking of the $B-L$ symmetry, giving rise to a seesaw mechanism of neutrino mass generation. The Lagrangian then reads
\begin{equation}
\mathscr{L}=\mathscr{L}_{SM}-V(H,S_{BL})-\left(\frac12\lambda_{N_i}S_{BL}\bar N_i^c N_i+Y_{ij}\bar\ell_i H^\dagger N_j+h.c.\right),
\end{equation}
where the kinetic terms of the new particles have been omitted and, without loss of generality, it is assumed that the $N_i$ mass matrix is diagonal after the spontaneous breaking of $B-L$. In the above equation, $H$ is the Standard Model Higgs doublet and $V(H,S_{BL})$ includes all terms in the scalar potential involving $S_{BL}$ (and possibly $H$ too):
\begin{equation}
V(H,\sbl)=\mu_{S}^2\sbl^\dagger\sbl-\lambda_{HS} (H^\dagger H)\,(\sbl^\dagger\sbl)+\frac{\lambda_S}{2}(\sbl^\dagger\sbl)^2.
\label{eq:vhs}
\end{equation} 
In the unitary gauge $H$ and $S_{BL}$ can be written as
\begin{align}
H &= \begin{pmatrix}0 \\ \frac{h+v_{EW}}{\sqrt 2}\end{pmatrix},\\
\sbl &= \frac{h_{BL}+\vbl}{\sqrt 2},
\end{align}
where $v_{EW}=246~\gev$ while $v_{BL}$ sets the $B-L$ breaking scale, which can be much higher than the electroweak scale. The scalars $h$ and $h_{BL}$ mix with each other via the $\lambda_{HS}$ term in equation (\ref{eq:vhs}), giving rise to two mass eigenstates, $H_1$ and $H_2$, defined by 
\begin{align}
H_1&= h\,\cos\theta+h_{BL}\,\sin\theta,\\
H_2&= -h\,\sin\theta+h_{BL}\,\cos\theta,
\end{align}
where $\theta$ is the mixing angle. To satisfy the LHC bounds on the properties of the Higgs boson \cite{Khachatryan:2014jba}, $\theta$ should be small. Hence, to a good approximation we can identify  $H_1$ with $h$ and $H_2$ with $h_{BL}$.  The scalar observed at the LHC with a mass of about  $125~\gev$ \cite{Aad:2015zhl} is then $H_1$. The mass of $H_2$ and $\theta$ are instead free parameters of the model. 

The $B-L$ gauge boson acquires a mass, $\mzbl$, after the spontaneous breaking of the $B-L$ symmetry by $S_{BL}$,
\begin{equation}
\mzbl=2g_{BL}v_{BL},
\end{equation}
being  $g_{BL}$ the $B-L$ gauge coupling constant. For simplicity, we assume a negligible kinetic mixing between $\ubl$ and $U(1)_Y$ so that  $\zbl$ and the SM $Z^0$ do not mix with each other. The mass of  $\zbl$ and the gauge coupling $\gbl$ can be constrained with collider data. From LEP II data  the bound 
\begin{equation}
\frac{\mzbl}{\gbl}\gtrsim 6-7~\tev
\end{equation}
was derived in \cite{Carena:2004xs,Cacciapaglia:2006pk}. At the LHC, current searches for dilepton resonances by ATLAS and CMS can also be used to set a bound on $\zbl$ through the Drell-Yan process ($q\bar q\to \zbl\to \ell\bar \ell$, with 
$\ell=e,\mu$) \cite{Chatrchyan:2012oaa}. This bound turns out to be sligthly more stringent than the LEP II bound for $\mzbl\gtrsim 3~\tev$ --see \cite{Guo:2015lxa}.    

After the breaking of the $B-L$ symmetry, the right-handed neutrinos also become massive, with $M_{N_i}=\lambda_{N_i}\vbl/\sqrt{2}$. (For simplicity, in our numerical analysis we take the right-handed neutrinos to be degenerate but our results hardly depend on this assumption). Thus, this model can naturally account for the observed pattern of neutrino masses and mixing angles via the seesaw mechanism.

This minimal $\ubl$ extension, however,  does not contain any dark matter candidates. We will therefore add one additional complex scalar field, $\phidm$, to play that role. $\phidm$ is a singlet of the Standard Model, has charge $n_{DM}$ under $\ubl$ and does not acquire a vacuum expectation value. As we show below, $n_{DM}$ can be easily chosen in such a way  that  it is not possible to write a  gauge invariant Lagrangian term that allows $\phidm$ to decay, rendering  $\phidm$  automatically stable. In other words, the $B-L$ symmetry itself  stabilizes the dark matter particle, making unnecessary an ad hoc discrete symmetry. Regarding the dark matter phenomenology, $\phidm$ will behave as a standard WIMP, obtaining its relic density via a freeze-out process in the early Universe, and giving rise to the usual  signals in direct or indirect detection experiments, as we will describe in the next section.

$\phidm$ has both gauge ($B-L$) and scalar interactions. The gauge interactions are determined, besides  $\gbl$ and $\mzbl$, by $\ndm$ and $\mdm$.  The scalar potential, on the other hand, now includes additional terms  involving the dark matter field:
\begin{align}
V(\phidm,H,\sbl)&=\mu_{DM}^2\phidm^\dagger\phidm+\lambda_{H}(\phidm^\dagger\phidm)(H^\dagger H)\nonumber\\
&+\lambda_{DM}(\phidm^\dagger\phidm)(\sbl^\dagger \sbl)+\lambda_4\left(\phidm^\dagger \phidm\right)^2,
\label{eq:v}
\end{align}
with $\mu_{DM}^2$, $\lambda_H$, $\lambda_{DM}$ and $\lambda_4$ free parameters. The dark matter mass, $\mdm$, is then given by 
\begin{equation}
\mdm^2=\mu_{DM}^2+\lambda_H\frac{v_{EW}^2}{2}+\lambda_{DM}\frac{\vbl^2}{2}>0.
\label{eq:mdm}
\end{equation} 
Notice that the dark matter mass, which is expected to be around the TeV scale to explain the observed dark matter density, receives a contribution proportional to $\vbl$. If $\vbl$ were much larger than the TeV scale, say $10^7~\gev$, an unnatural cancellation, or fine-tuning, between the first and third terms in equation (\ref{eq:mdm}) would be required to obtain the correct dark matter mass.  To avoid such excessive fine-tuning,  in our numerical analysis we take the dark matter mass as a free parameter but subject to the constraint  $\mdm> \sqrt{\lambda_{DM}}\vbl/3$. Even if somewhat arbitrary, this condition allows us to easily exclude the fine-tuned regions while scanning the parameter space of the model.  Thus, once we include scalar dark matter in the $B-L$ model, the breaking scale, $\vbl$, cannot be arbitrarily large and should lie relatively close to the TeV scale. In consequence, this setup gives rise to  a low scale realization of the seesaw mechanism.

Let us now discuss precisely what values can $n_{DM}$ take to ensure dark matter stability within this model. Terms that would allow dark matter to decay are either Yukawa interactions of the form $\phidm \bar f f'$ or the scalar interactions  $\phidm H_i$, $\phidm H_iH_j$ and $\phidm H_iH_jH_k$, where $H_i$ denotes either the SM Higgs doublet or $S_{BL}$.  To prevent dark matter decay we must therefore choose $n_{DM}$ in such a way that all these terms are forbidden. Regarding the Yukawa interactions, $\phidm \bar f f'$, since $\phidm$ is a singlet under the SM gauge group, the combination $\bar f f'$ should also be a  singlet of $SU(3)_c\times SU(2)_L\times U(1)_Y$. And the only  combination of fermions that  produces such a singlet is  $f=N^c_i$, $f'=N_i$, yielding a term  with $B-L=\pm 2$.  Thus, for $n_{DM}\neq \pm 2$ Yukawa interactions do not induce dark matter decay. Regarding scalar interactions, the mixing term $\phidm S_{BL}$ can be excluded with $n_{DM}\neq \pm 2$. Forbidding the cubic term $\phidm H^\dagger H$ requires $n_{DM}\neq 0$, which means that $\phidm$ cannot be a singlet of $B-L$. From the other possible cubic term,  $\phidm S_{BL}^2$, we get instead the condition $n_{DM}\neq\pm 4$. Similarly,  the quartic interaction $\phidm S_{BL}^3$ can be prevented with $n_{DM}\neq \pm 6$. Finally, one may even forbid decays via dimension-5 operators such as $\phidm S_{BL}^4$ (presumably suppressed by the Planck scale), which can be achieved for $n_{DM}\neq \pm 8$. Summarizing, the dark matter particle in this model will be stable provided that $n_{DM}\neq\pm 2n$, for $n\in \mathbb{Z}$ and $n\leq 4$.  Therefore, most choices of $n_{DM}$ actually lead to a stable dark matter particle.       

\section{Scalar Dark Matter}
\label{sec:dm}

In this section we study the dark matter phenomenology of this model, in particular the $\phidm$ relic density  and its detection prospects. Specifically, we determine the regions in the parameter space of the model where the dark matter constraint can be satisfied according to  the way in which the relic density is achieved in the early Universe. First, we will analyze the $B-L$ gauge interactions and later  the scalar interactions.

\subsection{Gauge interactions}
\begin{figure}[t!]
\begin{center}
\begin{tabular}{cc}
\includegraphics[width=0.4\textwidth]{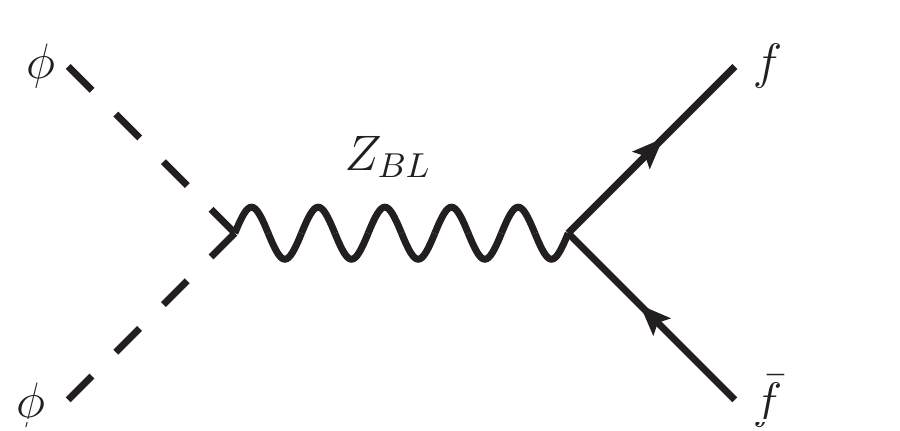}
& \includegraphics[width=0.4\textwidth]{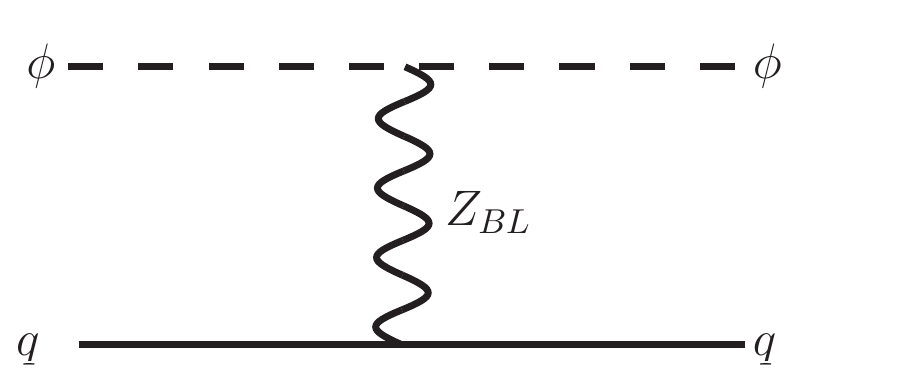}
\end{tabular} 
\caption{\small\it Left: The Feynman diagram corresponding to dark matter annihilation into fermions, $\phi\phi^\dagger\to f\bar{f}$, via the B-L gauge interaction. Right: The Feynman diagram for the spin-independent interaction between the dark matter and quarks.}
\label{feyngauge}
\end{center}
\end{figure}
Since $\phidm$ cannot be a singlet of $B-L$, it must necessarily have $\ubl$ gauge interactions. Let us then first consider the effects of these interactions  on the dark matter phenomenology. Surprisingly, these effects have not been considered before, likely as a result of the strong bounds on $\gbl$ and $\mzbl$. But as we will show, the $B-L$ gauge interactions may not only explain the relic density  but give also rise to a very predictive scenario; XENON1T should probe this possibility in the near future.

The gauge  interactions allow the annihilation of the dark matter particle into fermions mediated by the $B-L$ gauge boson, $\phi\phi^\dagger\to Z^*_{BL}\to f\bar{f}$ (see figure \ref{feyngauge}, left), as well as the direct annihilation into two gauge bosons, $\phi\phi^\dagger\to \zbl\zbl$.  The relevant parameters in this case are then just four: the dark matter mass ($\mdm$), the $B-L$ quantum number of $\phidm$ ($\ndm$), the $B-L$ gauge coupling ($\gbl$), and the mass of the $B-L$ gauge boson ($\mzbl$). As we will see, due to the strong experimental constraints on $\mzbl/\gbl$, only the annihilation into fermions turns out to be relevant, and only close to the $\zbl$ resonance.

\begin{figure}[t!]
\begin{center}
\includegraphics[width=0.8\textwidth]{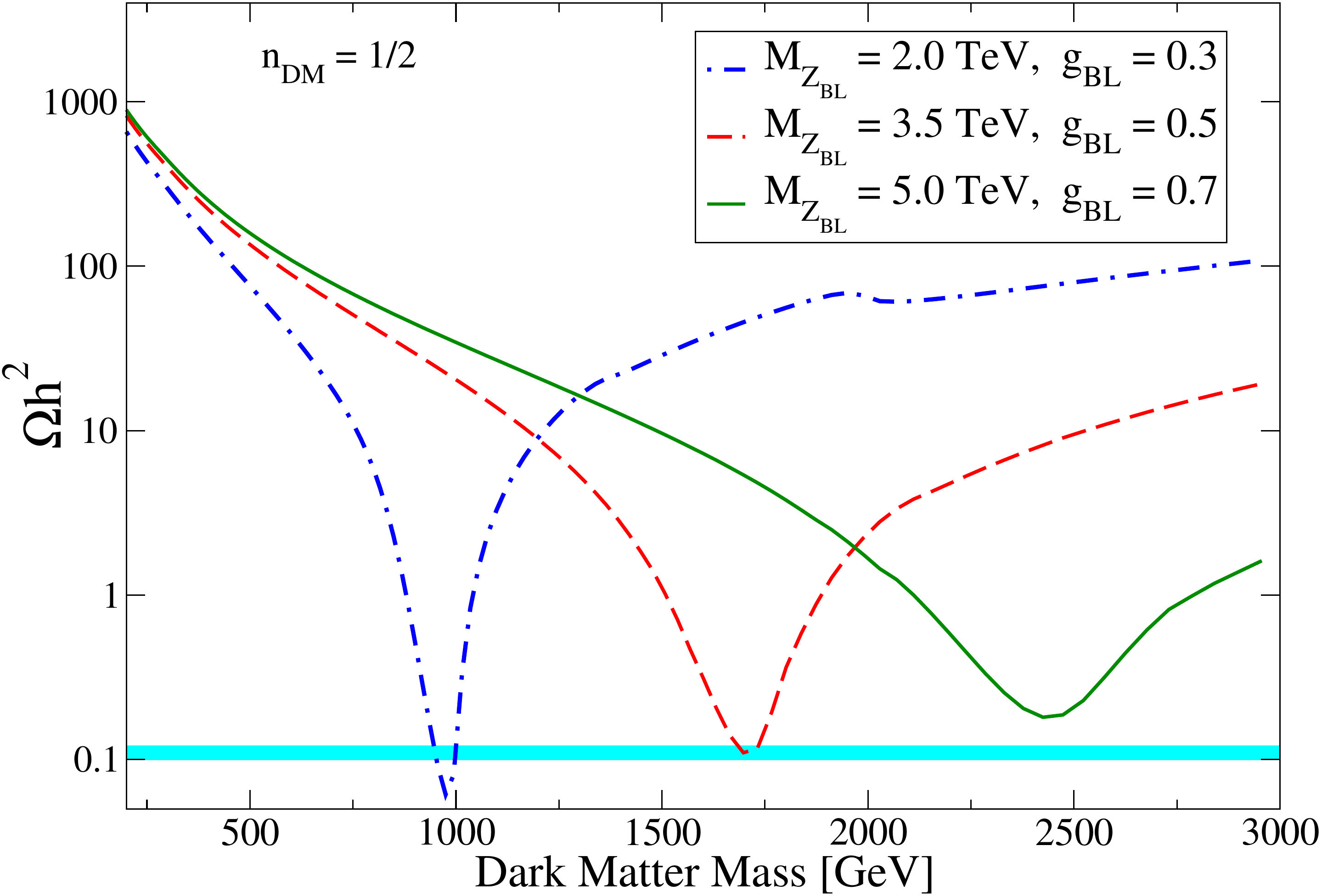} 
\caption{\small\it The dark matter relic density as a function of the dark matter mass for $\ndm=1/2$ and three different values  of ($\mzbl$, $\gbl$). In this figure it is assumed that the gauge interactions determine the relic density. The horizontal (cyan) band displays the region consistent with cosmological observations.}
\label{relicgauge}
\end{center}
\end{figure}

Dark matter annihilations are determined by the diagram shown in the left panel of figure 1. Due to the structure of the scalar gauge coupling, the resulting $\sigmav$ is velocity suppressed ($\propto v^2$). All our numerical results were obtained by implementing this model into micrOMEGAs \cite{Belanger:2013oya} via LanHEP \cite{Semenov:2010qt}. Figure \ref{relicgauge} shows the relic density as a function of the dark matter mass for $\ndm=1/2$ and three different values of ($\mzbl$, $\gbl$), all of them compatible with collider bounds. The behaviour is analogous in all three cases, with the relic density decreasing as $\mdm$ gets close to the resonance ($\mdm\sim \mzbl/2$), where it reaches its  minimum value, and increasing as the dark matter mass moves away from it.  Notice that the relic density tends to be much  higher than the observed value (the horizontal cyan band) except in a narrow region close to the resonance. And even at the resonance, the relic density can be too large to be in agreement with the data, as illustrated by the solid (green) line. That is, sitting at the resonance may not guarantee a sufficiently small relic density. For the dash-dotted line, the impact of the annihilation into $\zbl\zbl$ can be observed, as it gives rise to a slight suppression of the relic density once this channel becomes kinematically available ($\mdm\sim 2~\tev$), but it is too small to have a significant effect. That is why only the annihilation into fermions mediated by $\zbl$ is relevant in this case. From the figure it is clear  that, as expected, the higher the dark matter mass (or the gauge boson mass) the more difficult it is to satisfy the relic density constraint, and that for $\ndm=1/2$ it is not possible to satisfy it for dark matter masses above $2~\tev$ or so. Next, we will investigate exactly what this upper limit on $\mdm$ is and how it depends on $\ndm$. 

\begin{figure}[t!]
\begin{center}
\includegraphics[width=0.8\textwidth]{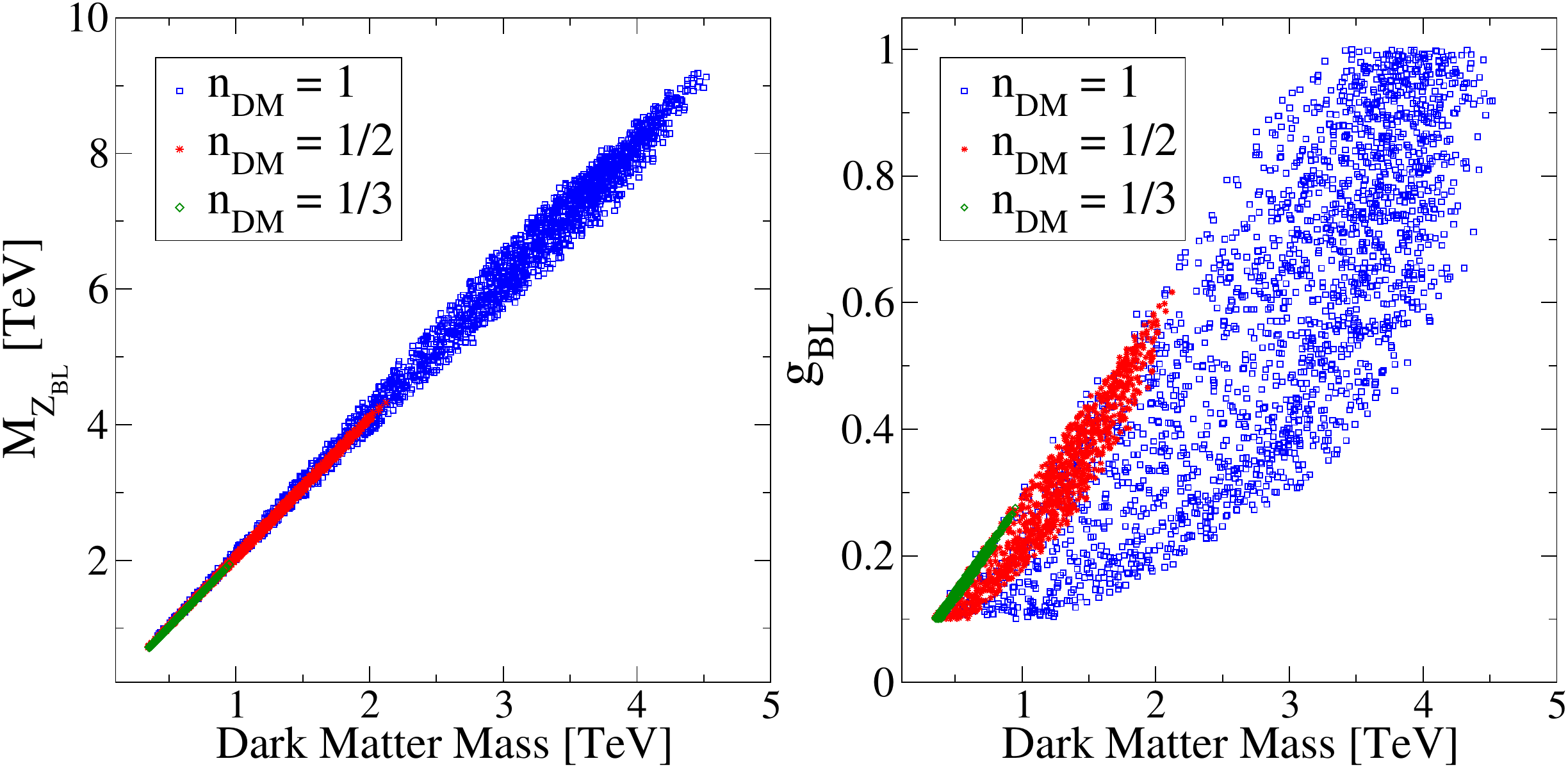} 
\caption{\small\it The viable parameter space  when the dark matter relic density is determined by the $B-L$ gauge interaction. Left: Viable points projected onto the plane ($\mdm$, $\mzbl$) for different values of $\ndm$: $1/3$ (green), $1/2$ (red), and $1$ (blue). Right: Viable points projected onto the plane ($\mdm$, $\gbl$) for different values of $\ndm$.}
\label{scangauge}
\end{center}
\end{figure}

To that end, we have randomly varied $\mdm$, $\gbl$ and $\mzbl$ for $\ndm=1/2,\,1/3,\,1$, and obtained a sample of viable models --those satisfying the relic density constraint,  collider bounds and the  perturbativity condition $\gbl<1$. In the following we will analyze such a sample. Figure \ref{scangauge} projects this set of  viable models onto the planes ($\mdm$, $\mzbl$) in the left panel and ($\mdm$, $\gbl$) in the right panel. The color convention is green, red and blue respectively for $\ndm=1/3,1/2,1$. From the left panel, we see that all the viable models indeed lie close to the resonance, with some non-negligible spread for $\ndm=1$. The range of dark matter masses increases with $\ndm$, extending from about $1$ TeV for $\ndm=1/3$, to $2$ TeV for $\ndm=1/2$, and to $4.5$ TeV for $\ndm=1$. From the right panel we see that larger values of $\gbl$ can be obtained  as $\ndm$ increases. They reach  $0.3$, $0.6$ and $1.0$ (the perturbative limit we imposed) respectively for $\ndm=1/3,1/2,1$. This figure also explains why  we find no viable models featuring $\mdm\gtrsim 4.5~\tev$: they  require  a gauge coupling larger than one --a possibility we did not consider.

\begin{figure}[t!]
\begin{center}
\includegraphics[width=0.8\textwidth]{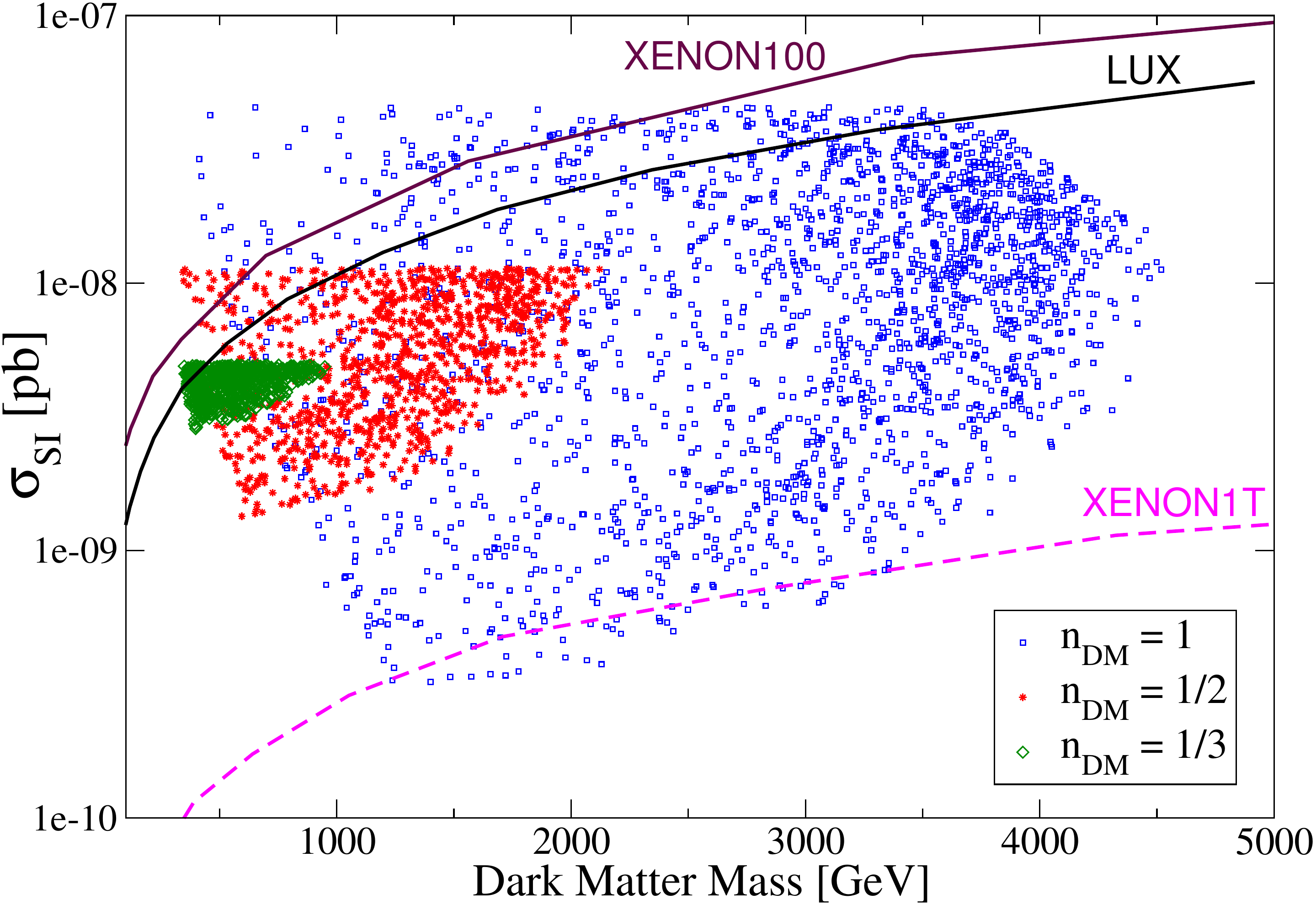} 
\caption{\small\it The spin-independent direct detection cross section as a function of the  dark matter mass for our set of viable points, classified according to the value of $\ndm$: $1/3$ (green), $1/2$ (red), and $1$ (blue). All these points satisfy the relic density constraint via gauge-mediated dark matter annihilations. For comparison, we also display the current bound from LUX (solid line) and the expected sensitivity of XENON1T (dashed line).}
\label{scangaugesip}
\end{center}
\end{figure}

The same $B-L$ gauge interaction that determines the relic density induces a coupling between the dark matter and quarks that can be probed in direct detection experiments --see figure \ref{feyngauge}, right. Being the dark matter candidate a scalar, only the spin-independent cross section, $\sip$, is relevant. Figure \ref{scangaugesip} displays  our sample of viable models in the plane ($\mdm$, $\sip$) using the same color convention as before. For comparison, the present bounds from XENON100 \cite{Aprile:2012nq} and LUX \cite{Akerib:2013tjd} (solid lines) are also shown, as well as the expected sensitivity of XENON1T \cite{Aprile:2012zx} (dashed line). When $\ndm=1/3$ (green points), the models lie in a tiny area featuring $\mdm\lesssim 1~\tev$ and $2.7\lesssim\sip/10^{-9}\mathrm{pb}\lesssim 5$ --very close to the LUX bound. A small fraction of them are in fact already excluded and the rest can be easily probed in future experiments.  When $\ndm=1/2$ (red points), the models lie in a larger area extending to dark matter masses of order $2~\tev$ and with cross sections between $10^{-8}$ and $10^{-9}$ pb. This region will be entirely probed by XENON1T. Finally, when $\ndm=1$ (blue points), the models occupy a broader area that extends down to cross sections of order $3\times 10^{-10}$ pb. Practically all of them lie within the expected sensitivity of XENON1T. And the few  lying below the XENON1T sensitivy  can be probed by either XENON-nT or LZ \cite{Malling:2011va}. Thus, this scenario can be tested in the near future via direct detection experiments.  

Regarding indirect detection, no signals are expected in this case because the annihilation rate today is velocity suppressed and therefore negligibly small. This fact could be used to falsify this scenario if a credible indirect detection signal of dark matter were observed.

Summarizing, dark matter annihilations mediated by the $B-L$ gauge boson allow to satisfy the relic density constrained but only close to the resonance ($2\mdm\sim \mzbl $) and within a narrow range of dark matter masses that depends on $\ndm$. Moreover the spin-independent direct detection cross section is expected  to be within the reach of the XENON1T experiment, providing a direct way to probe this scenario.

\subsection{Scalar interactions}
Besides the gauge interactions we have already studied, the dark matter particle, $\phidm$, also has scalar interactions with both $H\sim H_1$ (the Higgs) and $S_{BL}\sim H_2$ --see equation (\ref{eq:v}). Their effect on the dark matter phenomenology was partially studied in \cite{Guo:2015lxa}, but only within a scale invariant framework. Here we generalize their findings.

Scalar interactions between the dark matter and the SM scalar  give rise to the well-known Higgs-portal scenario \cite{Silveira:1985rk,McDonald:1993ex,Burgess:2000yq}, whose phenomenology has been thoroughly studied in the recent literature --see e.g. \cite{Yaguna:2008hd,Goudelis:2009zz,Guo:2010hq,Djouadi:2012zc,Cline:2013gha,Feng:2014vea}. Notice however that when the scalar Higgs-portal arises from the $B-L$ model, one automatically obtains  an explanation for the stability of the dark matter not based on an ad hoc $Z_2$ symmetry. This fact may be relevant in view of the possible effects due to dark matter decay induced by the breaking of discrete symmetries at the Planck scale, as recently emphasized in \cite{Mambrini:2015sia}. In addition, the dark matter field is necessarily complex --charged under $\ubl$-- rather than real, as usually assumed in standard Higgs-portal models. 

In the following we will focus on the interactions with $H_2$ or, to a good approximation, with $S_{BL}$, the scalar that breaks the $B-L$ symmetry. Such interactions give rise to two distinct final states in dark matter annihilations: $N_iN_i$ (mediated by $H_2$) and $H_2H_2$.  The relevant parameters in this case are the right-handed neutrino masses ($\mni$), the mass of $H_2$ ($\mhtwo$), the scalar coupling ($\ldm$), the vev of $S_{BL}$ ($v_{BL}$), and the dark matter mass ($\mdm$).

\begin{figure}[t!]
\begin{center}
\includegraphics[width=0.8\textwidth]{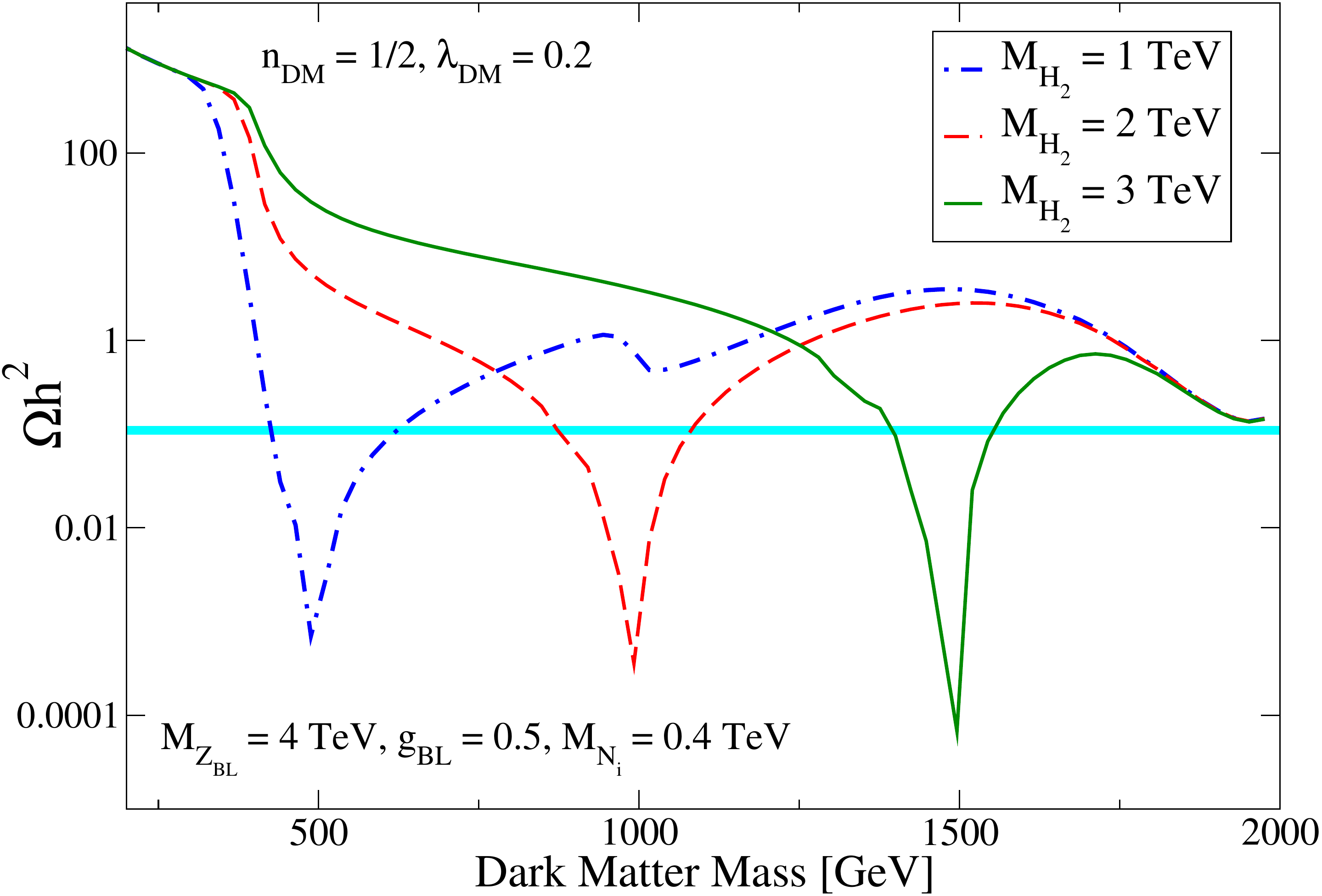} 
\caption{\small\it The dark matter relic density as a function of the dark matter mass for $\ldm=0.2$ and different values of $\mhtwo$: $1~\tev$ (dash-dotted line), $2~\tev$ (dashed line), and $3~\tev$ (solid line). The horizontal (cyan) band displays the region consistent with cosmological observations.}
\label{relicscalar}
\end{center}
\end{figure}

Figure \ref{relicscalar} shows the relic density as a function of the dark matter mass for $\ldm=0.2$ and different values of $\mhtwo$: $1~\tev$ (dash-dotted line), $2~\tev$ (dashed line), and $3~\tev$ (solid line). The common right-handed neutrino mass was taken to be $0.4~\tev$. The effect of the $H_2$ resonance is clearly observed in this figure as it leads to a strong suppression of the relic density for $\mdm\sim \mhtwo/2$, reaching values as low as $10^{-4}$.  In fact, for the chosen  value of $\ldm$, the relic density constraint (cyan horizontal band) can only be satisfied close to  the $H_2$ resonance.  The effect of the other annihilation channel, $H_2H_2$, can also be observed for $\mhtwo=1~\tev$ (dash-dotted line), as it induces a slight decrease in the relic density when $\mdm\gtrsim 1~\tev$.

\begin{figure}[t!]
\begin{center}
\includegraphics[width=0.8\textwidth]{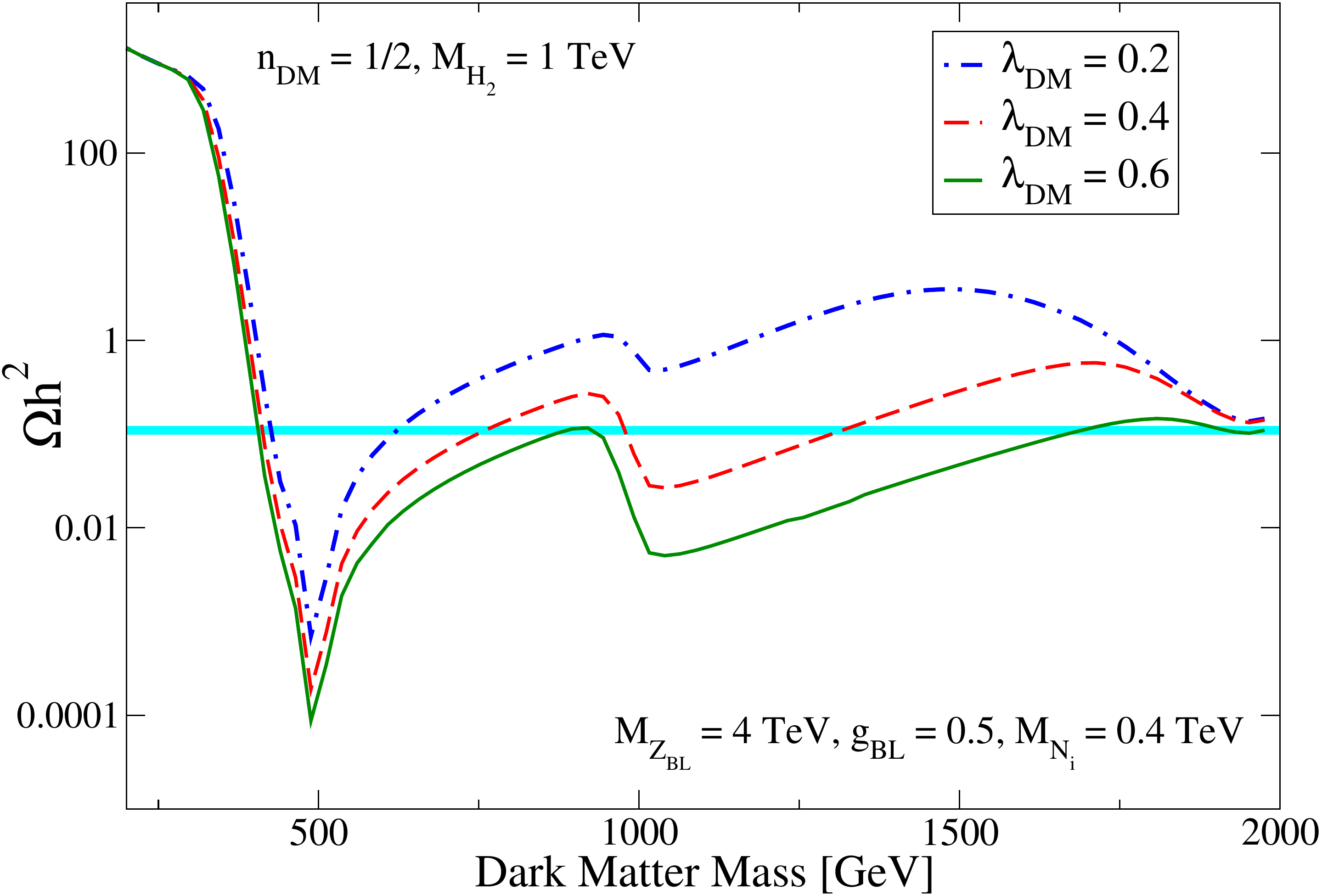} 
\caption{\small\it The dark matter relic density as a function of the dark matter mass for $\mhtwo=1~\tev$ and different values of $\ldm$: $0.2$ (dash-dotted line), $0.4$ (dashed line), and $0.6$ (solid line). The horizontal (cyan) band displays the region consistent with cosmological observations.}
\label{relicscalarla}
\end{center}
\end{figure}

To illustrate the dependence with $\ldm$, we show in figure \ref{relicscalarla} the relic density as a function of $\mdm$ for $\mhtwo=1~\tev$ and $\ldm=0.2$ (dash-dotted line), $0.4$ (dashed line), and $0.6$ (solid line). As expected, the larger the coupling, the smaller the relic density. A qualitative difference is that the larger couplings allow to satisfy the relic density constraint far from the $H_2$ resonance, via the annihilation into $H_2H_2$. For $\ldm=0.4$, the relic density is below the observed value for dark matter masses between $1~\tev$ and $1.3~\tev$ while for $\ldm=0.6$ this happens for $\mdm$ between $0.9~\tev$ and $1.7~\tev$. In the following we will study in more detail the resulting viable parameter space. To do so, it is convenient to distinguish between the two possible final states. First, we will address  dark matter annihilation into right-handed neutrinos ($\phi\phi^\dagger\to N_iN_i$), which, as we have seen, is relevant close to the $H_2$ resonance ($2\mdm\sim \mhtwo$). Second, we discuss dark matter annihilation into $H_2H_2$, which is possible  for $\mhtwo< \mdm$.

\begin{figure}[t!]
\begin{center}
\includegraphics[width=0.8\textwidth]{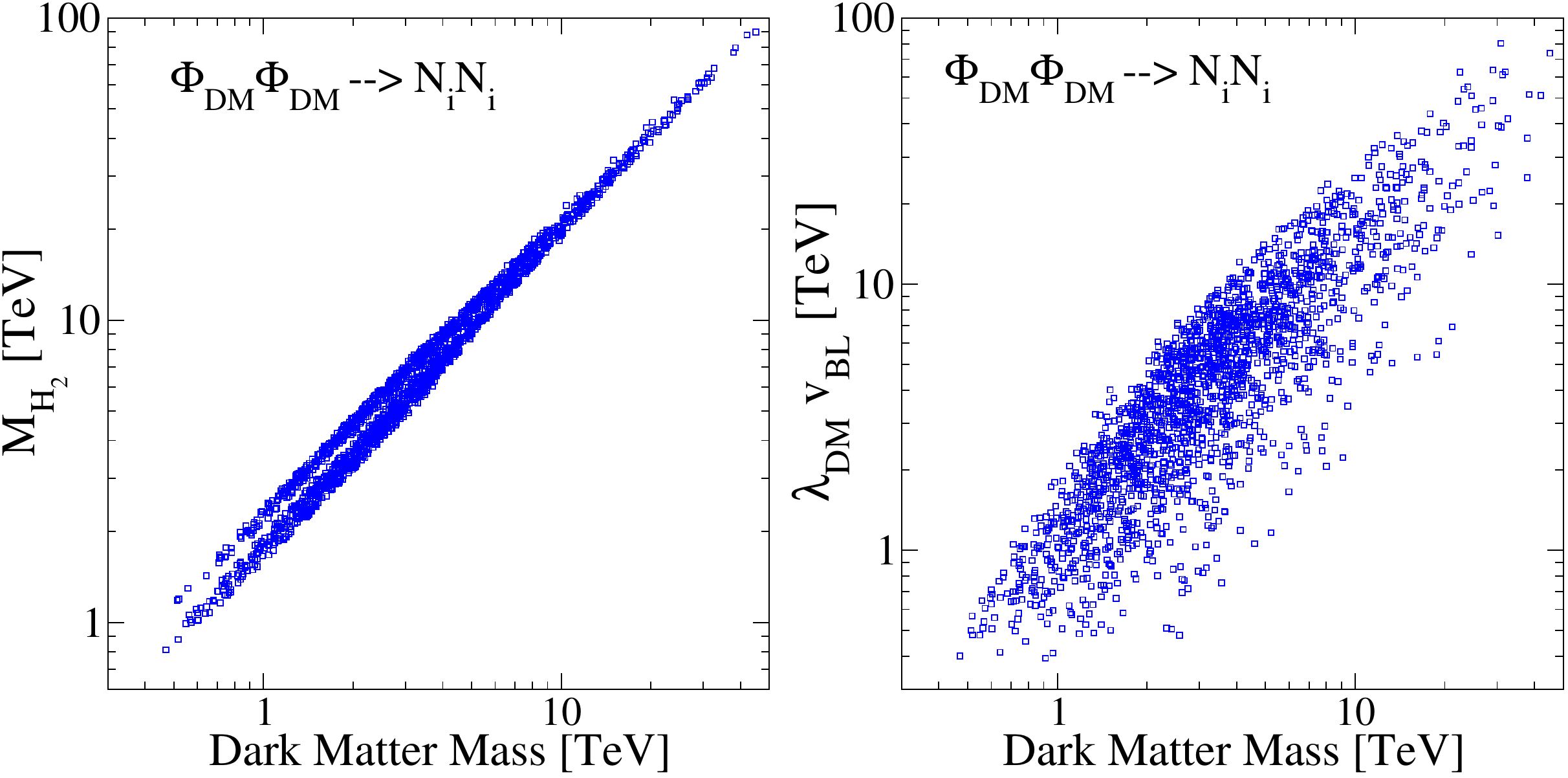} 
\caption{\small\it The viable parameter space  when the relic density is determined by dark matter annihilation into right-handed neutrinos --$\phidm\phidm^\dagger\to N_iN_i$. Left: The viable points projected onto the plane ($\mdm$, $\mhtwo$). Right: The viable points projected onto the plane ($\mdm$, $\ldm \vbl$).}
\label{scanscalar}
\end{center}
\end{figure}

Let us start with $\phidm\phidm^\dagger\to N_iN_i$, which proceeds through an $s$-channel diagram mediated by $H_2$, giving rise to a $\sigmav$ proportional to $(\ldm \vbl)^2$. Interestingly, this annihilation process involves three of the four new fields included in this $\ubl$ extension --only $\zbl$ is missing. In \cite{Guo:2015lxa} this process was briefly mentioned but not analyzed.  As before, we have scanned the relevant parameter space  and selected the points compatible with accelerator and perturbativity bounds, and with the dark matter constraint, requiring the relic density to be determined by the annihilation into right-handed neutrinos. The resulting viable models are shown in figure \ref{scanscalar}, projected onto the planes ($\mdm$, $\mhtwo$) in the left panel, and ($\mdm$, $\ldm \vbl$) in the right panel. From the left panel we see that for $\mdm\lesssim 2~\tev$ the viable models lie along two branches, one slightly above and the other slightly  below the $H_2$ resonance.  It is difficult to find viable models right on top of the resonance because the relic density tends to be quite suppressed there, as we already saw in figures \ref{relicscalar} and \ref{relicscalarla}. But as the dark matter increases, the two branches tend to merge and the viable models move closer and closer toward the resonance. At very high masses, the relic density constrained is satisfied only on top of the resonance.  Regarding the dark matter mass range, we see that, as a result of the resonant annihilation, it can extend to values as large as $50~\tev$. From the left panel we see that, as expected from the cross section,  the product $\ldm\,\vbl$ must increase with the dark matter mass in order to maintain the correct value of  the relic density, reaching, for $\mdm\sim 50~\tev$, a value of order $100~\tev$. For  $\vbl$ we find in our sample a maximum value of  about $200~\tev$. We see then how the dark matter constraint can set an upper bound on the otherwise unconstrained  $B-L$ breaking scale.

\begin{figure}[t!]
\begin{center}
\begin{tabular}{ccc}
\includegraphics[width=0.35\textwidth]{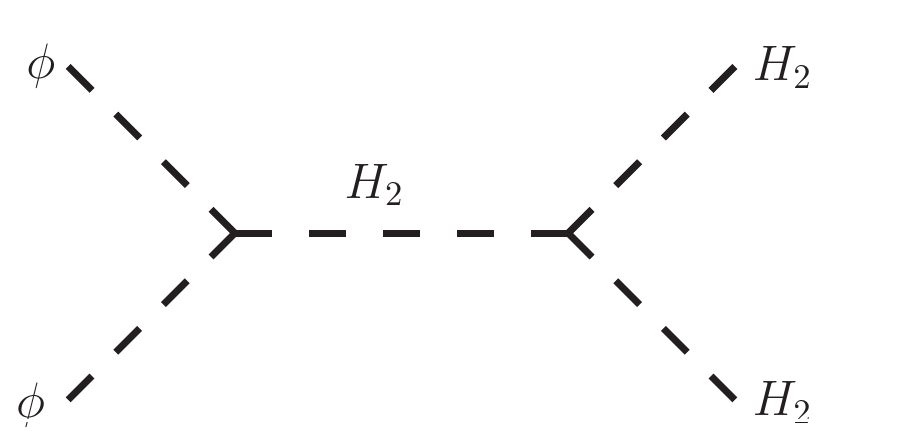}
& \includegraphics[width=0.35\textwidth]{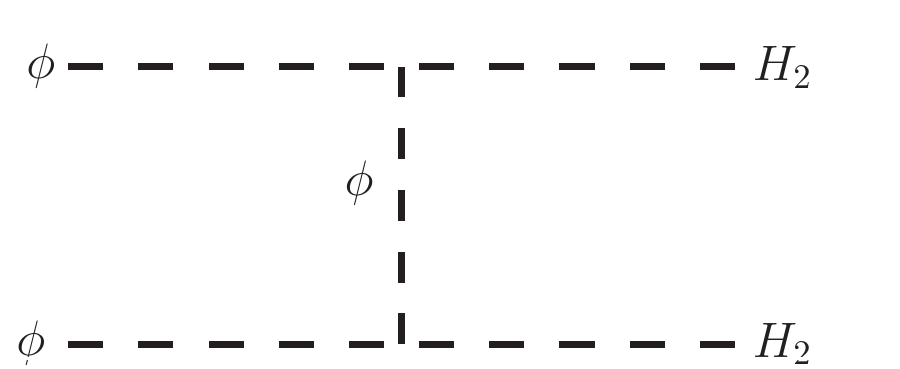} &\includegraphics[width=0.22\textwidth]{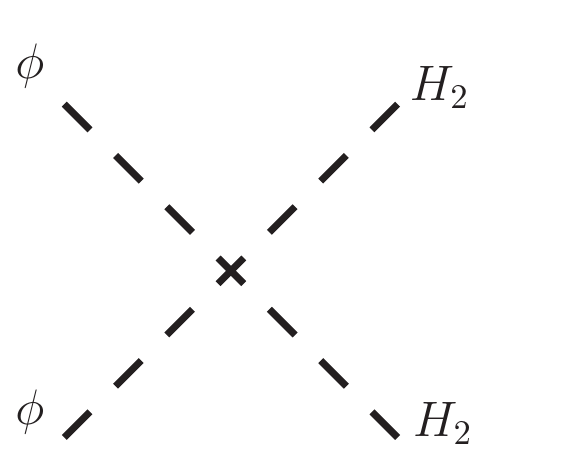}
\end{tabular} 
\caption{\small\it The Feynman diagrams that contribute to the annihilation of dark matter into the final state $H_2H_2$. }
\label{feynscah2}
\end{center}
\end{figure}

Now let us move to the second case, $\phidm\phi^\dagger_{DM}\to H_2H_2$. This is a non-resonant process requiring $\mhtwo<\mdm$ that can proceed through an $s$-channel  diagram mediated by $H_2$, a $t$- and $u$-channels mediated by $\phidm$, and via direct annihilation --see figure \ref{feynscah2}. This process was considered in detail in \cite{Guo:2015lxa}, where some analytical formulas can be found. Their numerical results, however, strongly depend on the assumed scale invariance and differ significantly from ours, as illustrated below. 

\begin{figure}[t!]
\begin{center}
\includegraphics[width=0.8\textwidth]{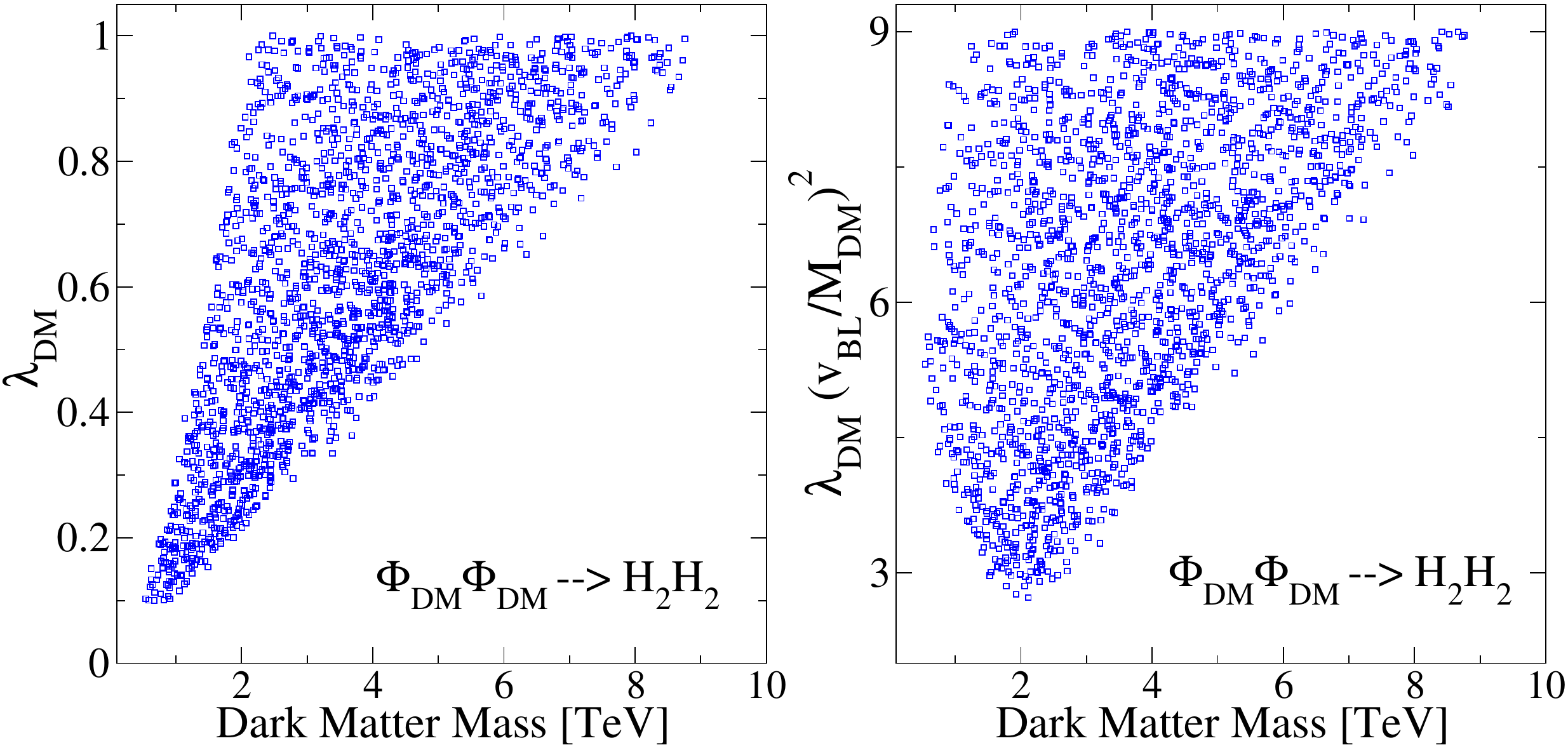} 
\caption{\small\it The viable parameter space  when the relic density is determined by dark matter annihilation into the heavier scalars --$\phidm\phidm^\dagger\to H_2H_2$. Left: The viable points projected onto the plane ($\mdm$, $\ldm$). Right: The viable points projected onto the plane ($\mdm$, $\ldm \vbl^2/\mdm^2$).}
\label{scanscalar2}
\end{center}
\end{figure}

We display the viable models in the plane ($\mdm$, $\ldm$) in the left panel of figure \ref{scanscalar2}. As expected, the minimum value of $\ldm$ increases with the dark matter mass so as to maintain the correct relic density. The maximum value of the dark matter mass in this case lies around $9~\tev$, where the observed dark matter density can be obtained only for $\ldm$ close to $1$ --the perturbative bound we have imposed. In the scale invariant version of this model, studied in \cite{Guo:2015lxa}, it was instead found that the relic density constraint could only be satisfied for $\ldm\gtrsim 1$. The right panel displays the viable models in the plane dark matter mass versus the product $\lambda_{DM} \vbl^2/\mdm^2$, which controls the relative contribution between the direct annihilation amplitude and the $t$- and $u$-channels --see \cite{Guo:2015lxa}. Since it is larger than $1$, it is the $t$- and $u$-channels that dominate the annihilation rate, and they are  more dominant at larger dark matter masses. In addition, that same product measures the fine-tuning required to obtain the dark matter mass, and we require it to be smaller than $9$ --see Section \ref{sec:mod}. From the figure we see that such an upper bound is often reached and that it helps set the upper limit on the dark matter mass.

The interaction between the dark matter and quarks proceeds in this case via diagrams mediated by $H_1$, $H_2$ and $\zbl$, which are not constrained by the relic density --they depend on additional parameters such as  $\theta$, $\lambda_{H}$, $\gbl$, and $\mzbl$. As a result, the direct detection cross section can vary over a very wide range. Regarding indirect detection, the advantage of the scalar interactions is that the annihilation rate is not velocity suppressed and it takes instead the expected value, $\sigma v\sim 3\times 10^{-26}\mathrm{cm^3s^{-1}}$. But since current dark matter indirect detection experiments (mostly those searching for gamma rays \cite{Ackermann:2013yva} and antiprotons \cite{Fornengo:2013xda}) only probe the small mass region ($M\lesssim 100~\gev$), no strong bounds can be obtained on this scenario.  

\subsection{Semi-annihilations}
The scalar interactions described in the previous section are generic, they are present for any value of $\ndm$. One can also have, however, interactions which are allowed  only for specific values of $\ndm$. For instance, when $\ndm=2/3$ the term\footnote{Another possibility is $n_{DM}=1$, which renders the term $\mu \phi^2_{DM}S_{BL}$ gauge invariant but it does not gives rise to semi-annihilations.}
\begin{equation}
\label{eq:semi}
\mathscr{L}_{2/3}=\frac{\lambda_{3}}{3}\phi_{DM}^3S_{BL}+h.c.  
\end{equation}
becomes compatible with the $B-L$ gauge symmetry. This term allows the dark matter particles to semi-annihilate \cite{D'Eramo:2010ep} via the process $\phi\phi\to\phi^\dagger H_2$ ($\phi\phi\to\phi^\dagger H_1$ also contributes but  is additionally suppressed by the mixing angle). For simplicity, we will assume in this section that all other interactions of the dark matter particle are irrelevant. Thus, the dark matter phenomenology is determined by only three parameters: $\lambda_3$, $\mdm$ and $\mhtwo$. Far from the $H_2$ threshold we get the simple behavior $\sigmav \propto \lambda_3^2/\mdm^2$. 

\begin{figure}[t!]
\begin{center}
\includegraphics[width=0.8\textwidth]{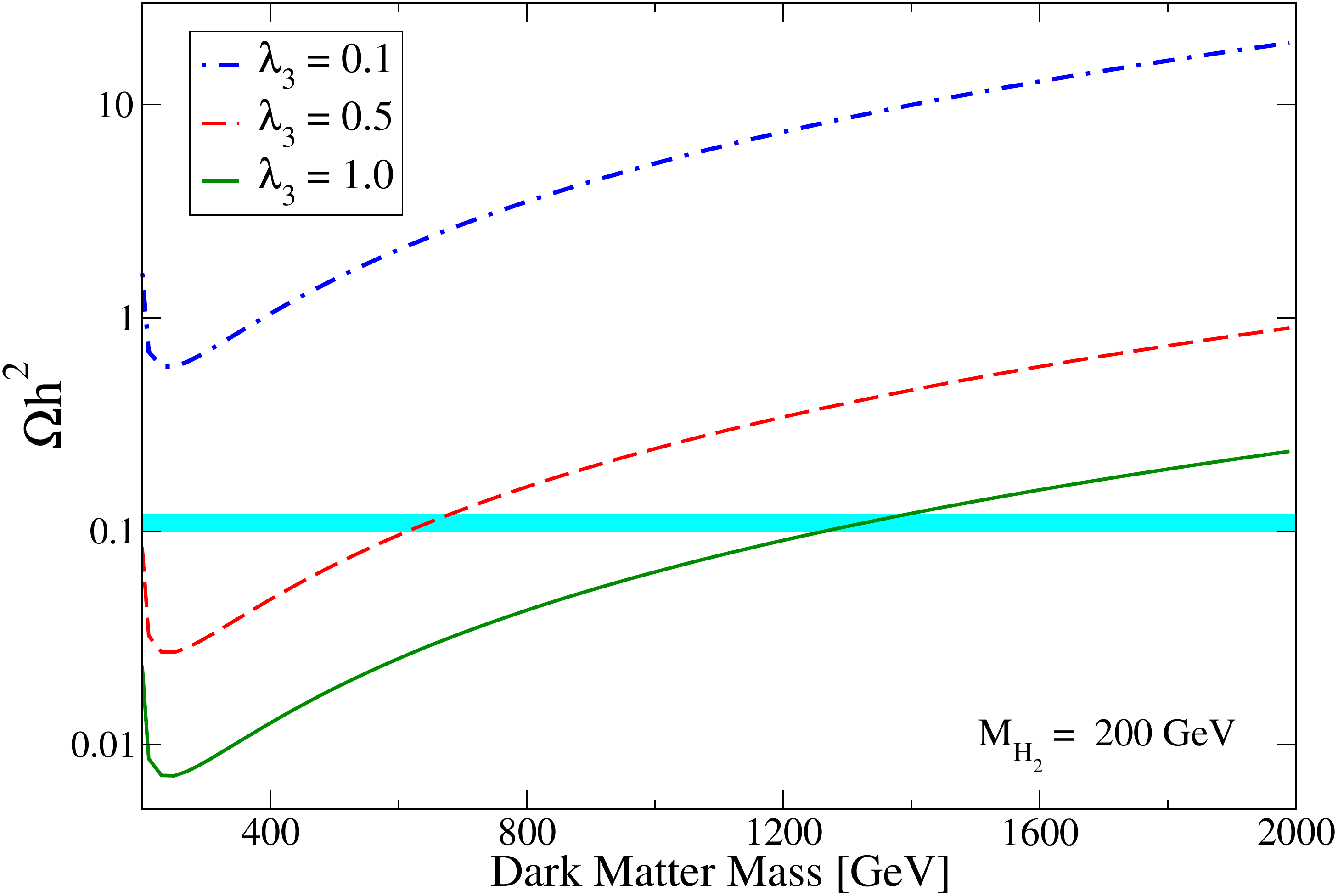} 
\caption{\small\it The dark matter relic density as a function of the dark matter mass for $\mhtwo=200~\gev$ and different values of $\lambda_3$: $0.1$ (dash-dotted line), $0.5$ (dashed line), $1.0$ (solid line). The horizontal (cyan) band displays the region consistent with cosmological observations. }
\label{relicsemi}
\end{center}
\end{figure}

Figure \ref{relicsemi} shows the  relic density as a function of the dark matter mass for $\mhtwo=200~\gev$ and $\lambda_3=0.1,\,0.5,\,1.0$. For comparison the region consistent with current observations is also displayed as a horizontal band. The behavior of the relic density for a given value of $\lambda_3$ is easy to understand: it simply increases with the square of the dark matter mass except close to the kinematic threshold ($\mdm\sim\mhtwo$). Notice that for $\lambda_3=0.1$, the resulting relic density is always above the observed range. For $\lambda_3=0.5$ and $\lambda_3=1$ compatibility with the observed data can be obtained respectively for $\mdm\sim 700~\gev$ and $\mdm\sim 1.3~\tev$. Since we require $\lambda_3<1$, we can already conclude that semi-annihilations do not allow to satisfy the relic density constraint for dark matter masses above $1.3~\tev$ or so.

\begin{figure}[t!]
\begin{center}
\includegraphics[width=0.8\textwidth]{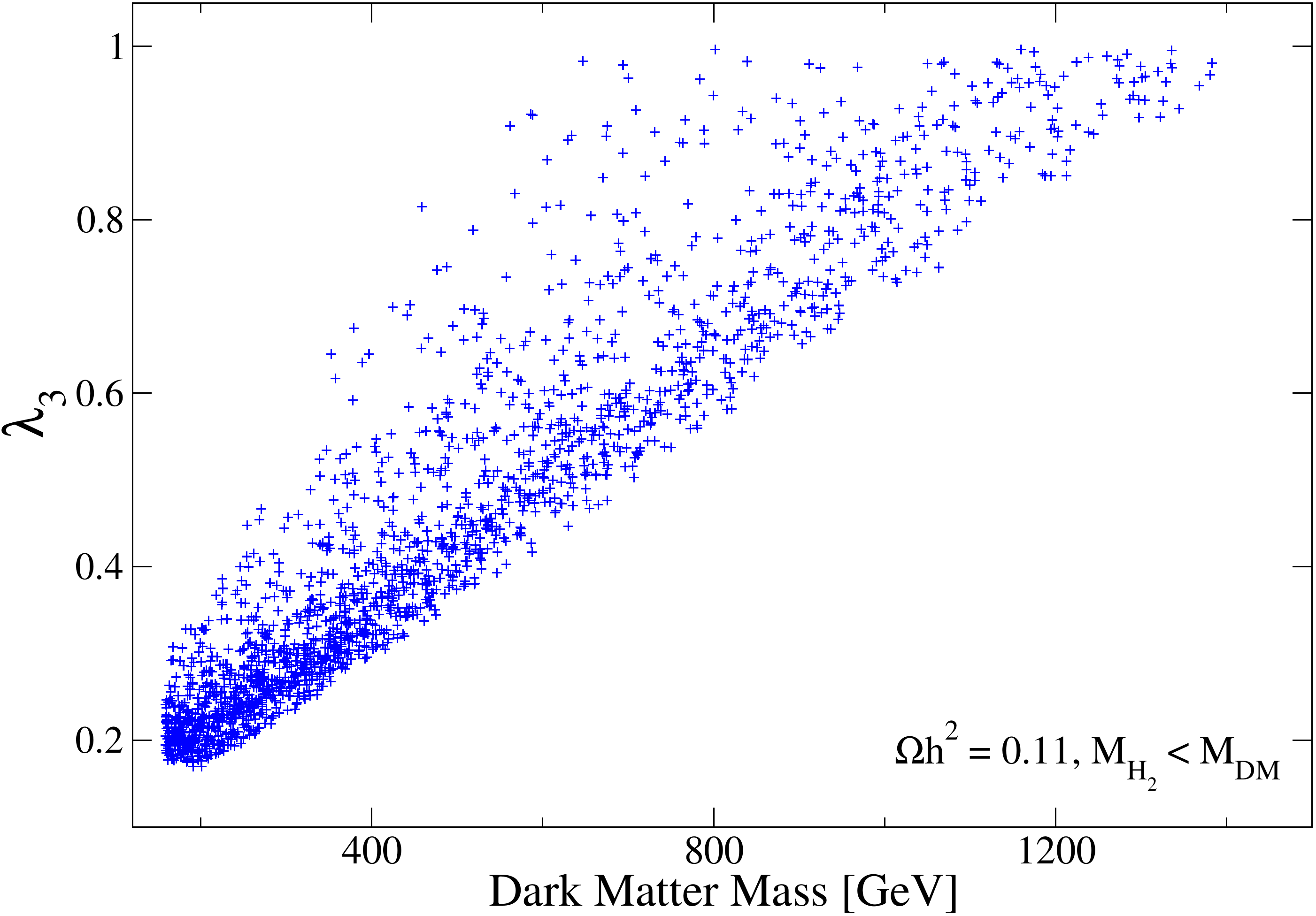} 
\caption{\small\it The viable parameter space  when the relic density is determined by semi-annihilations, projected onto the plane ($\mdm$, $\lambda_3$).}
\label{scansemi}
\end{center}
\end{figure}

This fact is further illustrated in figure \ref{scansemi}, which shows a scatter plot of  viable models in the plane ($\mdm$, $\lambda_3$). For this figure we assumed that semi-annihilations determine the relic density.  Since $\lambda_3$ should increase with the dark matter mass to keep the relic density constant, most points lie along  a band in this plane. The upper bound on the dark matter mass is obtained when  $\lambda_3=1$ and it reaches about $1.4~\tev$. 

Since the interaction in equation (\ref{eq:semi}) cannot give rise to dark-matter nucleon scattering, there is no prediction for the spin-independent direct detection cross section, which will be determined by the usual diagrams mediated by $H_{1,2}$ and $\zbl$. The annihilation rate today, on the other hand, does depend on $\lambda_3$. In fact, it was stated in \cite{Guo:2015lxa} that this semi-annihilation process could account for the  gamma ray excess from the Galactic center claimed in \cite{Goodenough:2009gk}. 

\section{Multi-component Dark Matter}
\label{sec:multi}

As we have seen in the previous sections, by simply  adding an extra scalar field charged under $B-L$, the $\ubl$ extension of the Standard Model  can account for the observed dark matter  in the Universe.  The stability of the dark matter is then guaranteed by the choice of the $B-L$ dark matter charge whereas the observed relic density can be obtained via annihilations induced by gauge or scalar interactions, or via semi-annihilations. It turns out that this minimal framework can also be extended in a straightforward way to a dark matter sector consisting of multiple dark matter particles, a scenario known as multi-component dark matter.

Many models, as a result of the discrete symmetry that is used to stabilize the dark matter, only allow for a single dark matter particle --typically the lightest odd state under a $Z_2$ symmetry. But there is currently no evidence that the dark matter sector is so simple. In principle, it could consists of a number of distinct particles with different properties, just like the visible sector.  Models with multi-component dark matter have certainly been studied before, particularly those with two-component  dark matter. And  they typically require  larger discrete symmetries, such as $Z_2\times Z'_2$ \cite{Aoki:2012ub,Bhattacharya:2013hva}, to stabilize the dark matter particles (an exception is the model considered in \cite{Esch:2014jpa}).  Within the $B-L$ model, multi-component dark matter  has the advantage of not requiring discrete symmetries at all and of allowing even for more than two dark matter particles.

Let us  qualitatively describe how this extension would work out for two dark matter particles, $\phi_1$ and $\phi_2$. To avoid mixing between them, their $B-L$ charges must be different, say $1/2$ and $1/3$. As before, this choice guarantees that both will automatically be stable without the need of an additional discrete symmetry. Regarding the relic density, there are essentially two different ways in which it can be obtained. The first one is a combination of the possibilities mentioned in the previous section. For instance, $\phi_1$ could annihilate resonantly via the gauge interaction whereas $\phi_2$ annihilates via scalar interactions into $H_2H_2$; or $\phi_1$ could annihilate instead into right-handed neutrinos at the $H_2$ resonance. The second way is via the annihilation of dark matter into dark matter; that is, the heavier dark matter particle annihilates into the lighter one ($\phi_2^\dagger\phi_2\to \phi_1^\dagger\phi_1$) via the new scalar interaction 
\beq
\mathscr{L}\subset \lambda_{12}\left(\phi_2^\dagger\phi_2\right)\left(\phi_1^\dagger\phi_1\right),
\eeq
which is always allowed. The lighter dark matter particle would then annihilate through one of the ways mentioned in the previous section. 

Even though this idea can in principle be extrapolated to any number of dark matter particles ($\phi_1,\ldots, \phi_n$), there is a limit on this number imposed by the dark matter constraint, which becomes stronger in models with multiple dark matter particles. As evidenced by the fact that it is often necessary to rely on resonances  to satisfy the dark matter bound, the scalar relic density tends to be high in this model.  If we now include several dark matter particles, the observed dark matter density would be the sum of the relic densities of all of them. Thus, each particle should contribute less than required in the previous section, making it  more difficult to satisfy the  dark matter constraint. But it is not hard to find viable regions for a small number of dark matter particles.

\section{Conclusions}
\label{sec:con}

We studied in detail scalar dark matter in the $B-L$ extension of the Standard Model. In it, the SM particle content  is extended by three right-handed neutrinos required to cancel the gauge anomalies, one scalar field to break the $B-L$ symmetry, and another scalar field charged under $B-L$ to explain the dark matter. This model naturally incorporates the seesaw mechanism of neutrino mass generation while  dark matter stability is easily achieved by appropiately choosing   the $B-L$ charge of $\phidm$. Moreover,  since the dark matter mass receives a contribution proportional to $\vbl$, the relic density constraint implies that the scale of $B-L$ breaking cannot be much above the TeV scale.  

The dark matter particle in this scenario has both gauge and Yukawa interactions and its relic density can be the result of annihilations or semi-annihilations. We analyzed these different possibilities and determined the viable parameter space in each case. If the dark matter relic density is determined by the  $B-L$ gauge interactions ($\phidm\phidm^\dagger\to \zbl\to f\bar f$), the annihilations should happen very close to the resonance ($\mdm\sim \mzbl/2$) and the dark matter mass cannot be larger than about $4.5~\tev$ (for $n_{DM}\lesssim 1$). In addition, the spin-independent cross section is expected to be large enough to be probed by the upcoming XENON1T experiment.  Scalar interactions, on the other hand, induce annihilations into right-handed neutrinos ($\phidm\phidm^\dagger\to H_2^*\to N_iN_i$) and into the heavier scalars ($\phidm\phidm\to H_2H_2$) via different channels. The former process takes place resonantly and allows the dark matter to be as large as $50~\tev$ whereas the latter one is non-resonant and requires a dark matter mass below $9~\tev$ to be in agreement with the observed  value of the dark matter density. When $n_{DM}=2/3$ semi-annihilation processes ($\phidm\phidm\to\phidm^\dagger H_2$) become possible and can account for relic density provided that $\mdm\lesssim 1.4~\tev$. Finally, we showed that one can easily accommodate multi-component dark matter in this scenario. The $\ubl$ extension of the Standard Model therefore provides a compelling and testable framework for neutrino masses and scalar dark matter.
\section*{Acknowledgments}
 We are supported by the Max Planck Society in the project MANITOP and by the DFG in the Heisenberg programme with grant RO 2516/6-1.

\bibliographystyle{hunsrt}
\bibliography{darkmatter}

\end{document}